\documentclass[]{raa}
\usepackage{graphicx,times}
\usepackage{natbib}
\usepackage{amssymb,amsmath}
\usepackage{rotating}
\usepackage{longtable}
\usepackage{array}
\usepackage{etoolbox}

\bibpunct{(}{)}{;}{a}{}{,}

\newcommand{\teff}{${T}_{\rm eff}$}
\newcommand{\logg}{log$g$}
\newcommand{\rc}{$r_c$}
\newcommand{\rr}{$r_0$}
\newcommand{\feh}{[Fe/H]}
\newcommand{\rv}{RV}
\newcommand{\kms}{km s$^{-1}$}
\newcommand{\rvz}{RV$_z$}

\newcommand{\rvmiu}{$0.85\pm1.26$}
\newcommand{\rvsig}{$5.47_{-0.71}^{+1.16}$}
\newcommand{\fehmiu}{$-0.08\pm0.04$}
\newcommand{\fehsig}{$0.13_{-0.02}^{+0.04}$}
\DeclareRobustCommand{\disambiguate}[3]{#2~#3}

\begin{document}

\title{Member candidates of the star clusters from LAMOST DR2 data }

\volnopage{ {\bf 2012} Vol.\ {\bf X} No. {\bf XX}, 000--000}
\setcounter{page}{1}

\author{Bo Zhang\inst{1,2}, Xiao-Yan Chen\inst{1}, Chao Liu\inst{1}, Li Chen\inst{3}, Li-Cai Deng\inst{1}, Jin-Liang Hou\inst{3}, Zheng-Yi Shao\inst{3}, Fan Yang\inst{1}, Yue Wu\inst{1}, Ming Yang\inst{1}, Yong Zhang\inst{4},Yong-Hui Hou\inst{4},Yue-Fei Wang\inst{4}
}


\institute{ Key Laboratory of Optical Astronomy, National Astronomical Observatories, CAS, Beijing 100012, China\ (NAOC); {\it liuchao@nao.cas.cn}\\
    \and
        University of Chinese Academy of Sciences, Beijing 100049, China\ (UCAS)\\
    \and
        Shanghai Astronomical Observatory, CAS, Shanghai 200030, China\ (SHAO)\\
    \and
        Nanjing Institute of Astronomical Optics $\rm \&$ Technology, National Astronomical Observatories, CAS, Nanjing 210042, China
\vs \no
{}
}
\date{Received~~2009 month day; accepted~~2009~~month day}

\abstract{In this work, we provide 2189 photometric- and kinematic-selected member candidates of 24 star clusters from the LAMOST DR2 catalog. We perform two-step membership identification: selection along the stellar track in the color-magnitude diagram, i.e., photometric identification, and the selection from the distribution of radial velocities, i.e. the kinematic identification. We find that the radial velocity from the LAMOST data are very helpful in the membership identification. The  mean probability of membership is  40\% for the radial velocity selected sample.  With these 24 star clusters, we investigate the performance of the radial velocity and metallicity estimated in the LAMOST pipeline. We find that the systematic offset in radial velocity and metallicity are \rvmiu\,\kms\ and \fehmiu\,dex, with dispersions of \rvsig\,\kms\ and \fehsig\,dex, respectively. Finally, we propose that the photometric member candidates of the clusters covered by the LAMOST footprints should be assigned higher priority so that more member stars can be observed.
}

\authorrunning{Zhang et al. }  
\titlerunning{Member candidates of star clusters}  
\maketitle

%
\section{INTRODUCTION}           
\label{sect:intro}
Star clusters are important constituents of galactic population and are natural laboratory of stellar evolution \citep{Oswalt2013} since their fundamental astrophysical parameters, e.g.  distance and age, could be determined more accurately than field stars. They are also used as the calibrators of the stellar astrophysical parameters for field stars in large surveys e.g. SDSS \citep{Lee2008,Xue2014} and LAMOST \citep{Xiang2015}. As a kind of important probes, they are often used to trace the global structure, kinematics, and evolution of the Milky Way \citep{Searle1978,Bica2003a,Chen2009,Frinchaboy2013}.

However, these advantages rely on efficient identification of the star clusters and their member stars.
Member star candidates can be roughly determined in photometry \citep[e.g., ][]{Bica2003a,Bica2003b,Dutra2003,Froebrich2007,Bukowiecki2011}.
The reliability of the memberships can be improved after including proper motion data in the identification \citep{Conrad2014,Kharchenko2012,Kharchenko2013}. Stellar parameters and radial velocities derived from the spectroscopic data can also be added to considerably improve the identification of the membership \citep{Frinchaboy2008,Kordopatis2013}. However, unlike the photometry and proper motion data, the sampling of the spectra is usually very low and thus the identification of the member stars of clusters may not be efficient in the spectroscopic surveys.

The sampling issue is now partly solved by the LAMOST survey, which has internally released about 3.8 million stellar spectra, making it the largest database of the stellar spectra in the world. The LAMOST telescope can simultaneously observe 200 objects per square degree, which is a factor of 2 higher than SDSS. Thanks to its high efficiency, LAMOST will essentially completely observe the stars brighter than $r=14$\,mag in the covered sky area of $\sim20000$ square degrees. The sampling rate for stars brighter than $r=17.8$\,mag is also high, compared with other surveys. On the other hand, LAMOST provides the radial velocity with accuracy at about 5\,\kms and measures [Fe/H] with accuracy better than 0.2\,dex \citep{Luo2015,Gao2015}. The high sampling rate with the accurate stellar parameter estimates is helpful for increasing the efficiency of the identification of the cluster members. However, because the limit of the mechanic design of the focal plane, LAMOST cannot concentrate more than 3 fibers within a small region as the typical core radius(a few arcmin) of a star cluster. Hence, although its survey footprints have covered $\sim457$ star clusters by now, it is very hard to perform a complete targeting for any individual cluster.

Even though the completeness of the membership of these hundreds of star cluters is low, it is still fairly valuable in the sense of 1) the calibration for both the fundamental parameters of the star clusters and the stellar parameters and distance estimates for the field stars observed by LAMOST survey \citep{Xiang2015,Carlin2015} and 2) mapping the chemo-dynamical evolution of the Milky Way \citep{Hou2013}.

For these purposes, we present the technique of the identification of the membership of clusters by combining the photometric with the LAMOST spectroscopic information.

The paper is organized as below. In Sect.~\ref{sect:data} we specify the data used in this work. In Sect.~\ref{sect:method} we describe the details of our method to determine the probability of membership. In Sect.~\ref{sect:results} we present the results. Discussions are raised in Sect.~\ref{sect:disc} and conclusions are drawn in Sect.~\ref{sect:summary}.

\section{DATA}\label{sect:data}
\subsection{LAMOST Spectroscopic Survey}
LAMOST telescope, also called Guo Shou Jing telescope, is a 4-meter reflected Schmidt telescope with a 5-degree field of view, on which 4000 fibers are installed. The spectral resolution is about $R=1800$ covering all optical wavelength \citep{Cui2012,Zhao2012}. The LAMOST survey contains both the LAMOST ExtraGAlactic Survey (LEGAS) and the LAMOST Experiment for Galactic Understanding and Exploration survey  \citep[LEGUE, ][]{Deng2012,Smith2012,XWLiu2014}.

The targets of LAMOST survey are selected from several catalogs, including UCAC4 \citep{Zacharias2013}, Pan-STARRS 1 \citep{Tonry2012}, Xuyi Galactic anti-center survey \citep{Yuan2015}, and 2MASS \citep{Skrutskie2006}. The apparent magnitude covers from $r=10$ to 17.8 \,mag. The photometric input catalog can be used as the  complementary data to present the observed star cluster in the color-magnitude diagram.

The LAMOST DR2 data, which includes observations from autumn 2011 to summer 2014, are used in this work. The complete photometric data covering each plate observed by LAMOST are also used to construct the color-magnitude diagram for the star clusters.

\subsection{The LAMOST Stellar Parameter Pipeline (LASP)}
The LAMOST DR2 catalog contains 4,136,482 targets, including 3,784,461 stars, 37,206 galaxies, 8,630 QSOs, and 306,185 unknown objects. For over half of the stars (2,207,788), the \emph{LASP} has successfully extracted stellar parameters, i.e. \feh, \logg, \teff, and radial velocity (\rv), from their spectra.

The core of \emph{LASP} is the ULySS algorithm \citep{Wu2011b}. To optimize the solutions and shorten the running time, an initial guess is firstly estimated by correlation function initial (CFI) method in the pipeline (Luo et al in prep.). ULySS starts with the initial guess and iteratively searches for the optimized stellar parameters in two parallel procedures with the non-normalized and the normalized (pesudo-continuum removed) spectra, respectively. The derived parameters with small difference between the non-normalized and normalized spectra are accepted.
Meanwhile, only high $g$ band signal-to-noise ratio (SNR) spectra of A-type stars(SNR$ > 40$), and F-, G-, and K-type stars (SNR$ > 6$ for dark night and SNR$ >15$ for bright night) are processed by \emph{LASP}. The typical error for \teff, \logg, \feh, and \rv\ are 140 K, 0.22\,dex, 0.12\,dex and 5\,\kms respectively \citep{Gao2015}.

\subsection{Star Cluster List}
We adopt the MWSC (Milky Way Star Cluster) catalog \citep{Kharchenko2012,Kharchenko2013} as the target list of star clusters since it provides homogeneous parameters of Milky Way star clusters and is complete in the observed volume of LAMOST. Thus we use the MWSC radius parameters for star clusters, i.e., $r_0$ in MWSC is the angular radius of the core of the cluster, and $r_2$ (hereafter rewritten as \rc) stands for the angular radius of the cluster. A star cluster is covered by LAMOST survey if the number of the LAMOST observed stars located within 2\rc\ of the cluster is larger than zero. In total, 457 star clusters, including open clusters, globular clusters, stellar associations, and moving groups, are found being covered by the LAMOST DR2 data.

\section{Method}\label{sect:method}

For each star cluster, the identification of membership is two-fold. First, we select the stars closed to the stellar track in the color-magnitude diagram. Then, we improve the identification in the distribution of the radial velocity. After the identification, we check out the reliability of the candidates
from position and proper motion diagram.

\begin{figure}[htbp]
    \centering
    \includegraphics[width=8cm, angle=0]{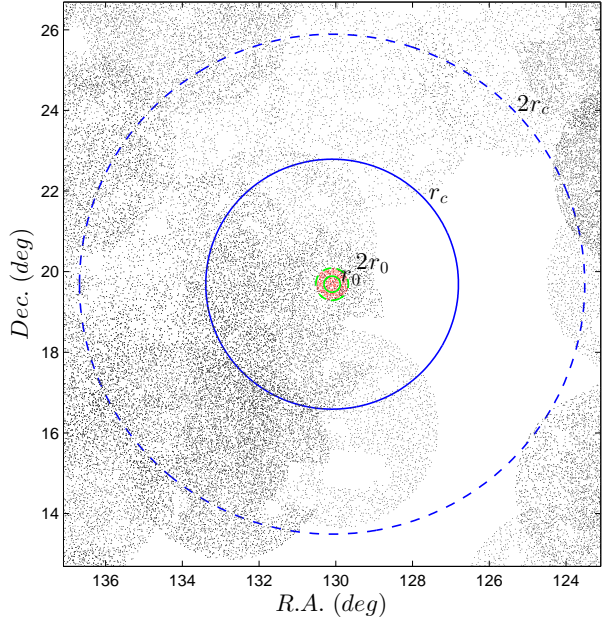}
    \caption{The sky atlas around the sample cluster NGC2632 (M44). The green (blue) solid line circle represent \rr\ (\rc, definitions are in text). Each dashed line circle has a radius of twice that of the corresponding solid line circle. The red dots indicate UCAC4 / Pan-STARRS1 data in green dashed line circle (within 2\rr) and black dots denote stars observed by LAMOST. The red dots in the green dashed circle are selected to highlight the intrinsic stellar locus of the cluster in the subsequent procedures (the red dots in Fig.~\ref{fig:cmd}).}
    \label{fig:radec}
\end{figure}

\subsection{Selection in color-magnitude diagram}
\begin{figure}[htbp]
    \centering
    \includegraphics[width=14.5cm, angle=0]{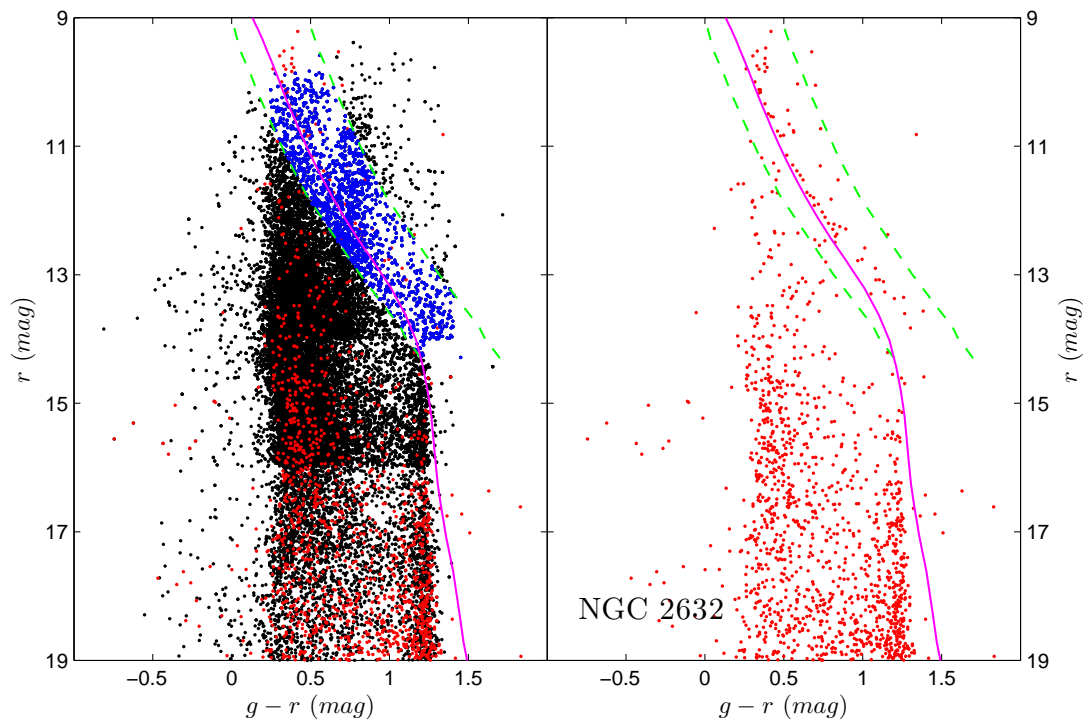}
    \caption{The sample $g-r$ vs. $r$ diagram for NGC2632. The black dots in the left panel are the stars observed by LAMOST(same as black dots in the dashed blue circle in Fig \ref{fig:radec}) within 2\rc, while the red dots in the panels are the stars within 2\rr. The pink line indicates the PARSEC synthetic stellar track  with the same metallicity, age, extinction, and distance of the cluster. The green dashed lines represent the boundaries used for candidate selection around the stellar track. They are manually set for each cluster and the typical width of the cut is $\sim$0.5\,mag in $g-r$. The blue dots in the left panel are the LAMOST observed stars located in both the 2\rc\ circle and the band around the stellar track defined by the green dashed lines.}
    \label{fig:cmd}
\end{figure}

We map all the stars observed by LAMOST located within 2\rc\ (the dashed blue circle in Fig~\ref{fig:radec}) on the $g-r$ vs. $r$ diagram in the left panel of Fig~\ref{fig:cmd} and the UCAC4 / Pan-STARRS1 data within 2\rr\ (the dashed green circle in Fig~\ref{fig:radec}) in the right panel taking NGC 2632 (M44) as an example. Fig~2$~$4 for other star clusters can be found in appendix.

It is noted that the samples show hard cuts at $r$=14, 16, and 19\,mag, which is because the selection function of the targets truncates the data at these positions for bright, middle, and faint plates, which are assigned short, intermediate, and long exposure time, respectively \citep{Yuan2015}. LAMOST observes the bright and middle plates mostly in bright / gray nights, but observes the faint plates only in clear dark nights. Hence, the bright and middle plates have more opportunity to be observed than the faint ones. Therefore, the target selection function prefer more brighter stars than the fainter ones \citep[ for more details, see][]{Yuan2015}. However, this does not means that, in a specific line of sight, the bright stars are always more frequently observed than the faint ones. The weather and the survey strategy may also alter the distribution of the $r$ magnitude in the observed samples. The region around NGC 2632, for instance, does not gain more bright plate observations. We point out that the specific selection effect function in the LAMOST input catalog would not significantly affect the identification of the stellar tracks of the star clusters in the color-magnitude diagram, since there are sufficient stars located within the core radii of the star clusters helping to lock down the stellar tracks. Indeed, it can be clearly seen in the right panel of Fig~\ref{fig:cmd},  in which the stars coming from the central part of the cluster show a prominent stellar track.

The stellar track is strengthened by overlapping a PARSEC synthetic stellar track \citep{Bressan2012} with same metallicity, age, interstellar extinction, and distance as NGC 2632. Stars located within 2\rc\ and  width of $\sim$0.5\,mag in $g-r$ around the stellar track are selected as the photometric member candidates. For those star clusters without MWSC \feh\ value, we assume \feh\ is 0 (solar-like) and select regions by according to photometry data to avoid the misleading by the problematic \feh. However, the turning point magnitude, the distance and log$t$ are still information for us. In most cases, 2\rc\ is a quite large region compared to the core radius of the star clusters. We use such a large radius to select the candidate because 1) the LAMOST fibers are very difficult to be multiply assigned in a very small region as mentioned before; 2) the membership in the outskirt region of the star clusters are particularly of interest which sheds light of the dynamical evolution of the stellar systems. Note that, after this procedure the peculiar stars, e.g. the blue stragglers and binary stars may be lost, which is a price we have to pay for the improvement of the efficiency of member star selection. We then apply this procedure to most of the covered 457 star clusters except those with prominent \rv\ peak and could be directly processed in next step. These special star clusters are marked using 'n' in the column 'CMD' in Table~\ref{tab:MWSC28p}

\subsection{Further selection in \rv\ distribution}\label{subsect:rv}
In most cases, member stars of a star cluster belong to a simple stellar population\footnote{Although recent studies find that some massive globular clusters contain multiple stellar populations \citep{deGrijs2010}, it would not significantly affect our technique since the age and metallicity of the multiple populations are only slightly different.}, i.e. they have same age, metallicity, and distance with very small dispersion. Moreover, the member stars show bulk motion with velocity dispersion of a few \kms\ and very concentrated proper motions. These characteristics allow us to identify the memberships by searching the clump of the stars in the space of the parameters. In general, we can map the candidate stars selected from the color-magnitude diagram to the multi-dimensional space of radial velocity--metallicity--proper motions and look for the clump. However, in practice, the measurement uncertainty of these observables are not equal. The radial velocity is usually measured in the highest accuracy in LAMOST data, and the error is about 5 \kms, which is slightly larger than the internal dispersions of the star clusters. The uncertainty of the metallicity measured from the LAMOST spectra is about 0.1-0.2\,dex, but quite close to the dispersion of the field disk stars.

\begin{figure}[htbp]
    \centering
    \includegraphics[width=10cm, angle=0]{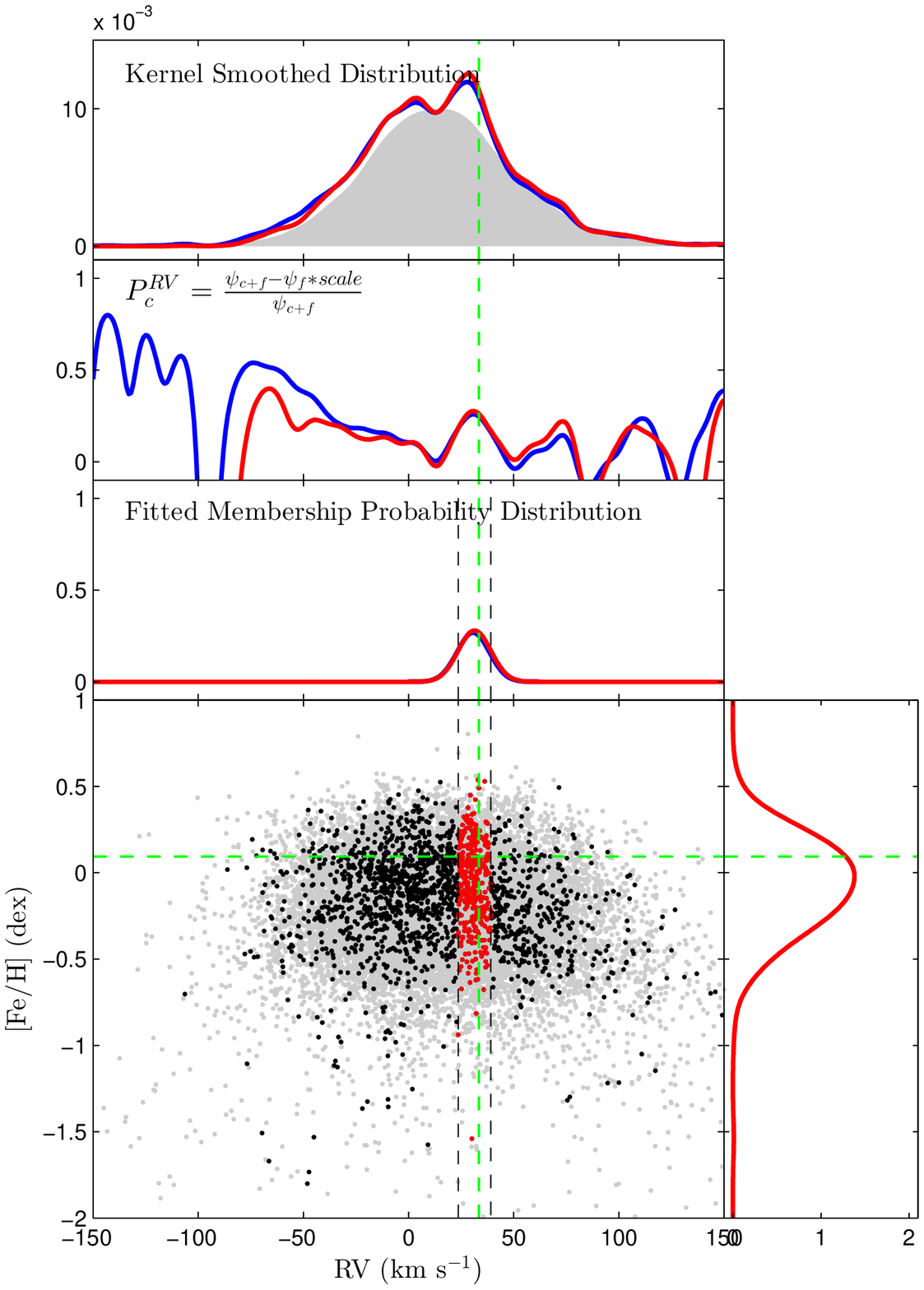}
    \caption{
    The demonstration of the kinematic member identification for NGC 2632. In the top panel, the red (blue) line denotes the Kernel-Smoothed distribution of the \rv\ (\rvz) of the CMD-selected stars, while the gray area denotes the scaled Kernel-Smoothed distribution of the \rv\ of stars located between 2 and 3 \rc\. The second panel shows the derived membership probability and the Gaussian-Fitted membership probability distribution.
    In the bottom panel, the gray and black dots are stars located within 2\rc\ within center of the cluster with and without CMD selection.
    The member star candidates selected using the distribution of \rv\ are turned into red.
    And the right panel shows the Kernel-Smoothed distribution of \feh\ of the candidates (red dots in bottom panel), the Gaussian Kernel bandwidth is 0.12 dex (typical error)}
    \label{fig:rvfeh}
\end{figure}

Despite the $\sim$2\,mas yr$^{-1}$ systematic error in PPMXL catalog, the 4$\sim$8\,mas yr$^{-1}$ random errors of the proper motions \citep{WuZY2011} may totally smear out the signature of the star cluster when the distance of the cluster is larger than a few hundreds parsecs.
Therefore, for our available data, the RV distribution is the best discriminator of cluster membership due to its smallest relative measurement errors.
We then use \rv\ for the further candidate selection and use  proper motions as the diagnostic to assess the performance of the selection in \rv. Fig.~\ref{fig:rvfeh} shows the sample distributions of \rv\ and \feh\ again with the cluster NGC 2632.
Since the LAMOST stellar parameter pipeline provides the accuracy of radial velocity estimates within 5\,\kms\ \citep{Gao2015}, larger than the intrinsic velocity dispersion of a star cluster, thus the measured velocity dispersion of member candidates of a cluster is dominated by the measurement error.

It is noted that there are two radial velocities released in the LAMOST catalog. One is from ULySS, which simultaneously derive the radial velocity with other stellar astrophysical parameters \citep{Wu2011a,Wu2011b,Wu2014}.  The other is measured from the cross-correlation with the ELODIE library (Luo et al in prep.). In general, the former one should be more accurate since the quality of the spectra involved in ULySS is higher, while spectra with any signal-to-noise ratio are handled in the ELODIE-based cross-correlation. Second, ULySS derives the radial velocity with relatively accurate stellar parameters, hence the template spectra used to find the radial velocity may well match the observed one, while the template spectra used in cross-correlation may not match the observed spectra due to the mis-classification. In this work, we intend to find as many as possible member candidates. Therefore, in order not to miss the spectra without stellar parameters, we adopt both radial velocities, the ULySS derived radial velocity, denoted as \rv, which is more accurate but only half of the spectra in the DR2 catalog have this value estimated; and the cross-correlation derived radial velocities, denoted as \rvz, which are less accurate but estimated for each spectrum. Thus, we divide the selected results into two versions, one includes stars with stellar parameters \rv, and the other includes all stars with \rvz.

In order to assign a membership probability to each star, we adopt the method proposed by \citep{Frinchaboy2008} but modified some details. Again, we use NGC 2632 as the sample to show how to purify the member candidates with the peak in \rv\ (or \rvz) distribution.
The lines in top panel of Fig~\ref{fig:rvfeh} is the normalized Kernel-Smoothed RV distribution (red for \rv, blue for \rvz) smoothed by a Gaussian Kernel with 5\,\kms\ bandwidth, which is the typical error of RV. Furthermore, we assume that stars outside 2\,\rc\ are 'non-members'  \citep{Frinchaboy2008}.
Thus we adopt the normalized Kernel-Kmoothed RV distribution of the stars located between 2\,\rc\ and 3\,\rc\ to represent the distribution of field stars ($\psi_f$). The gray area for Kernel-Smoothed RV distribution of stars located between 2 and 3 \rc\ scaled by a factor in order to fit the wings of the peak stacked by member stars.
To calculate the membership probability alongside the RV tick,
\citet{Frinchaboy2008} uses:
\begin{equation}\label{mpeq0}
  P_c^{RV}=\frac{\psi_{c+f}-\psi_{f}}{\psi_{c+f}},
\end{equation}
which will lead to significant negative value of the derived membership probability around the peak contributed by member stars and the wings of the peak of membership probability extends to negative when both $\psi_{c+f}$ and $\psi_{f}$ are normalized.
Moreover, this leads to underestimation of the membership probability of stars.
Thus we modified this equation by multiply the $\psi_f$ by a scale factor to solve this problem, and the new equation we adopt is
\begin{equation}\label{mpeq0}
  P_c^{RV}=\frac{\psi_{c+f}-\psi_{f}*scale}{\psi_{c+f}}.
\end{equation}
To calculate the scale factor for given $\psi_{c+f}$ and $\psi_{f}$ (this can not be done automatically since the scale factor is largely affected by the RV position of the star cluster), we specify an interval, e.g., [15.0 50.0]\,\kms\ for NGC 2632 (since there is a prominent peak around 35\,\kms), then we scale the $\psi_{f}$ to fit $\psi_{c+f}$ in [5.0 10.0]\,\kms\ and [50.0 55.0]\,\kms range (5\,\kms\ around both sides of the specified interval), and use least square to obtain the scale factor of the best fit. The fit range we specify for each star cluster and the scale factor obtained (Scale$_RV$ and Scale$_{RV_z}$) can be found in Table~\ref{tab:MWSC28p}.
The scale factor helps us to substract the field star component more accurately.

The next step is to fit the derived membership probability curve and extract the membership probability profile. We apply a local Gaussian Fit to the curve in our specified interval (e.g., [15.0 50.0]\,\kms\ for NGC 2632). The fit results are shown in the third panel of of Fig~\ref{fig:rvfeh}, with which we are able to assign a membership probability for each star given RV.
In order to estimate the bulk probability, in our catalog of member stars we only maintain the stars with relatively high membership probability (located within the interval [$center-\sigma,center+\sigma$], where $center$ and $\sigma$ are the of center position and dispersion of the fitted Gaussian Function, which is shown in the third and bottom panel of Fig~\ref{fig:rvfeh} using black dashed lines).
In the bottom panel of Fig~\ref{fig:rvfeh}, the black (gray) dots are the RV and \feh\ of all the stars located within 2\rc\ with (without) a CMD selection. The stars selected using RV method above and included in our member star candidate catalog are turned into red.

The right panel of Fig~\ref{fig:rvfeh} is the Kernel-Smoothed (by a Gaussian Kernel with bandwidth of 0.12\,dex) distribution of the \feh\ of our selected member star candidates.
Although, \feh\ can be used in the member identification, it may not significantly improve the performance since the most of the open clusters have very similar metallicity as the field stars and the measurement error of 0.1-0.2\,dex is much larger than the intrinsic dispersion. However, it is still very helpful to double-check the performance of the photometric-kinematic member identification. In the right panel, the distribution of \feh~for all stars is shown in gray and the distribution for kinematic member candidates in red. It is seen that, although it is hard to identify the member stars from the distribution of all stars, the kinematic member candidates do show a significant peak at around the green horizontal line, which is the metallicity of the cluster from the MWSC catalog.

\section{Results}\label{sect:results}

For each cluster, we identify the kinematic member candidates according to the technique mentioned in Sect~\ref{sect:method}. We leave the clusters with at least 3 kinematic member candidates based on \rvz\ in Table~\ref{tab:MWSC28p}, which includes 21 open cluster, 2 globular cluster, and 1 open cluster with nebulosity. In total, we identify 2189 member candidates from the Kernel-Smoothed distribution of \rv\ and 3559 from \rvz\ belonging to the star clusters and list them in Table~\ref{tab:table_member_rv}. The probability of the members in the candidate samples vary from 9\,\% to 100\,\% and from 7\,\% to 95\,\% for candidates selected based on \rv\ and \rvz. The overall probability is 40\,\% for both candidates selected based on \rv\ and \rvz. The catalog of the candidates selected based on \rv\ can be found in the appendix and those selected based on \rvz\ will be available as online material.

We take the median of the \rv\ and \feh\ of member star candidates for each star cluster as the estimated value of the bulk properties of the star clusters, and its distance to 85 and 15 percentiles (for normal distribution, the 15 and 85 percentiles are roughly $center-\sigma$ and $center+\sigma$) as the upper and lower error of the estimated value.
Table~\ref{tab:MWSC28p} lists all these LAMOST derived bulk parameters with a $\ast$ marked on the column names. Table~\ref{tab:table_member_rv} lists all the obtained member star candidates and Fig~2$\sim$4 for each clusters can be found in the appendix as well.

For the first time, we determine \rv\ for 4 of the 24 clusters and the \feh\ for 11 of them. Their parameters are marked with $\dag$ in Table\ref{tab:MWSC28p}. Compared with the large area coverage of the LAMOST survey, the number of the star clusters that have the identified member candidates in the LAMOST catalog is still very small. Therefore, we propose to mark higher priority to the photometric member candidates of the all 457 star clusters covered by LAMOST footprints in the input catalog, such that they can have higher opportunities to be observed in the rest years of the survey.

\section{Discussions}\label{sect:disc}

\subsection{Other parameters}
\begin{figure}[htbp]
    \centering
    \includegraphics[width=10cm, angle=0]{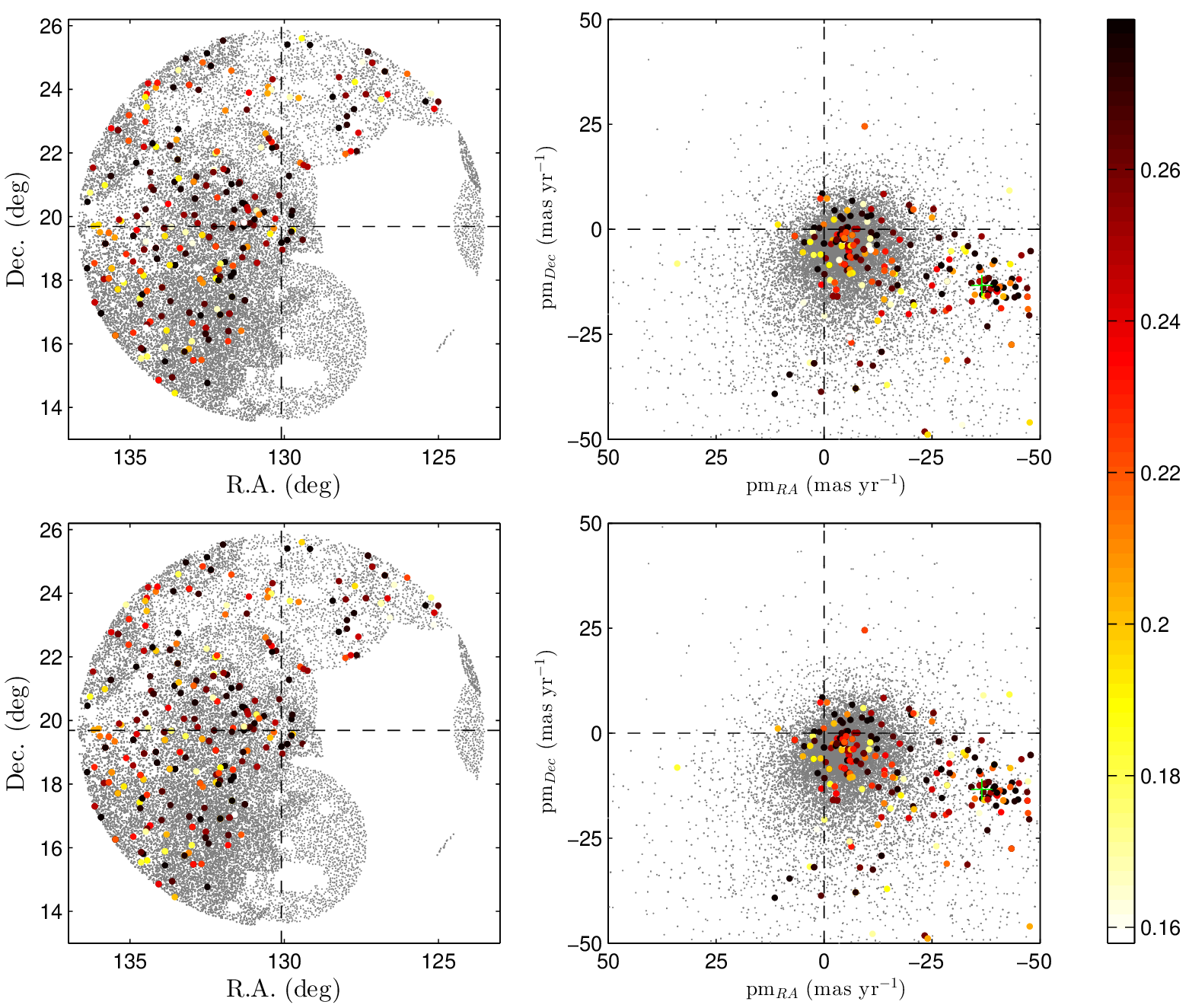}
    \caption{The spatial distributions (left panels) and proper motion distributions (right panels) for all stars inside 2\rc\ (gray dots) and the kinematic member candidates (colored dots in the top panels are member candidates based on \rv\ and those in the bottom panels are based on \rvz, for NGC 2632).
    The color of dots denotes the membership probability assigned to stars using our method.
    The black dashed lines indicate the central position of the cluster in left panels and the zero points of proper motions in the right panels.
    The green crosses in right panels denote the MWSC value of the bulk proper motion.}
    \label{fig:pmra}
\end{figure}

For each star observed with LAMOST, we find its best counterpart in the PPMXL catalog \citep{Roeser2010} within 3 arcsecs. If there are multiple objects from PPMXL within 3 arcsecs, we choose the nearest one as the counterpart. For some clusters, e.g. NGC 2632 (see Fig.~\ref{fig:pmra}), the proper motions are significantly different from that of the field stars, thus the proper motions can be very helpful in purifying the member candidates. However, for most clusters, the proper motions are too blur and flooded in the field stars and the large measurement errors as inferred in \citet{WuZY2011}. Consequently, they may not significantly improve the membership identification for most of our 24 star clusters.
We leave the further identification of the members with proper motions to the future users based on their specific requirements.
Moreover, not all LAMOST spectra have reliable proper motion measurements and the cross-identification may lead to significant decreasing of the number of the member candidates if we apply proper motions to the membership identification.

\begin{figure}[htbp]
    \centering
    \includegraphics[width=14.5cm, angle=0]{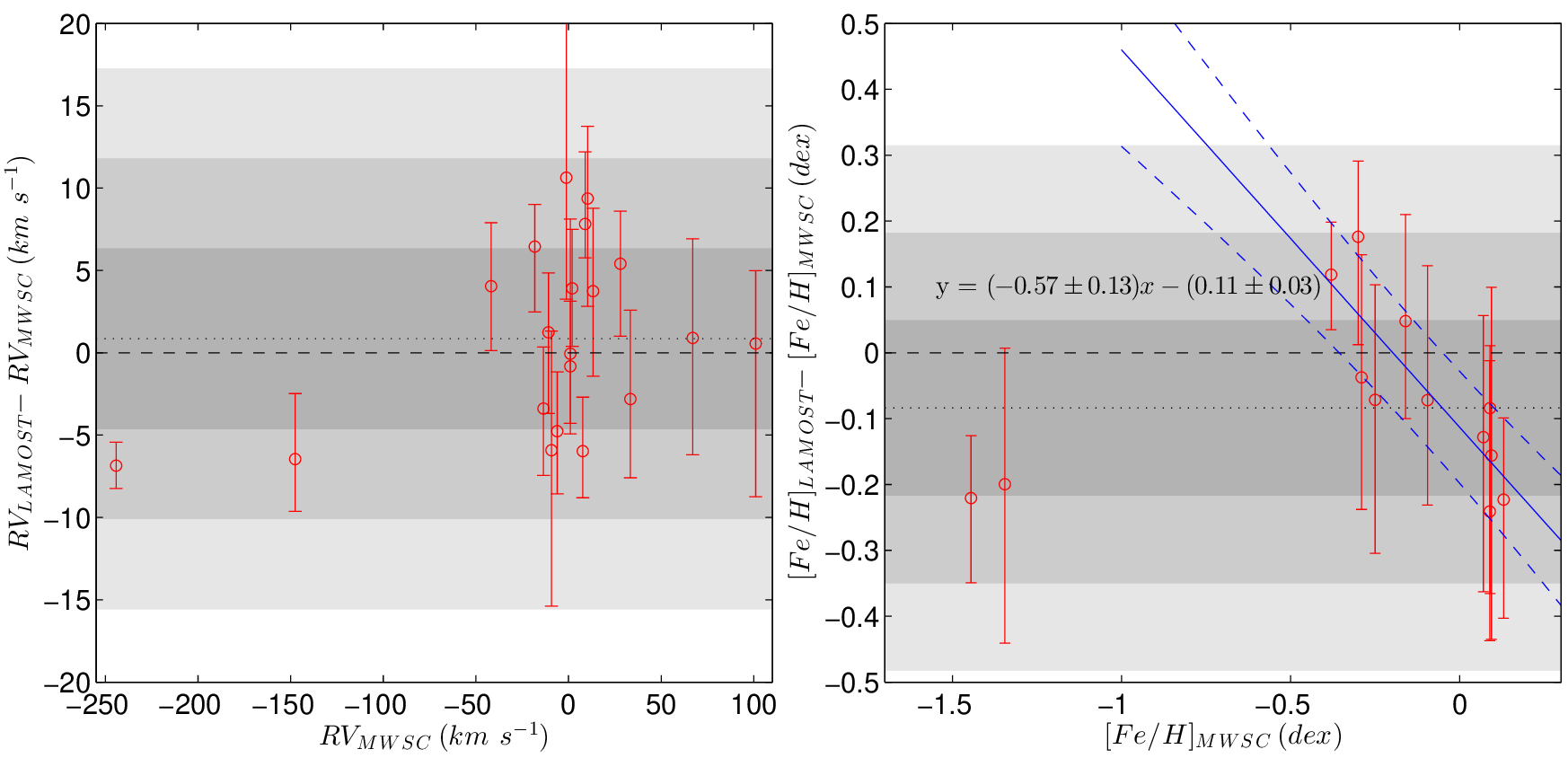}
    \caption{The residuals of \rv\ (left panel) and \feh\ (right panel) to the MWSC catalog. The red circles stand for the median values of the kinematic member candidates derived from the LASP for 24 clusters. The standard deviations of the members for each cluster are shown as the error bars. The black dashed line denotes the zero point and black dashed dot line indicates the systematic offset of \rv\ and \feh\ in the left and right panel, respectively. The area filled with gray color with different levels indicate 1, 2 and 3 $\sigma$ confidence interval from inside to outside.}
    \label{fig:rvcompare}
\end{figure}
\normalsize

\subsection{Caveat}
The membership probability is determined by the ratio of the number of members and field stars.
This value is not comparable with those in MWSC because the different definitions of the membership probability and the different types of data used for identifications.
The membership probability in MWSC catalog derived from Flux-Limited photometry surveys has a very simple selection effect while the membership probability derived from LAMOST RV data has a very different and more complicated selection effect, which is determined by the LAMOST targeting plan.
The value of our membership probability is defined by the fraction of the fibers pointing to the real members.
Consequently this probability is dominated by the efficiency of the LAMOST targeting on star cluster members.

\subsection{Comparison of the LAMOST derived \rv\ and \feh\ with MWSC catalog}

With the member candidates and their parameters obtained, we are able to compare the RV and [Fe/H] values of the cluster with those in the literature.
Adopting the MWSC parameters as their true values, we clearly see an overall consistency from Fig.~\ref{fig:rvcompare}. Using normal distribution statistics, the systematic offset of \rv\ is \rvmiu\,\kms, with a dispersion of \rvsig\,\kms, while for metallicity, the systematic offset and dispersion are \fehmiu\,dex and \fehsig\,dex, respectively.
Note that neither \rv\ nor \feh from MWSC catalog are used for membership identification. We do not find the systematic offset in \rv\ as mentioned in \citep{Luo2015,Gao2015,Xiang2015}, although a similar systematic offset is found in \feh. Moreover, the dispersion in \rv\ is well consistent with \citet{Gao2014}. And the dispersion in \feh\ is consistent with \citet{Gao2015}, which is 0.11 dex, as well.

For most star clusters, RV is very near the MWSC value except a few biased by $\sim1\sigma$, which is still within the tolerance.
However, [Fe/H] measurement is a bit worse than RV. For 2 metal-poor Globular Clusters, [Fe/H] estimated by LASP are significantly lower, and also the same case in the metal-rich end. This could be due to the template incompleteness or some bias in the pipeline. The parameters for these star clusters enable us to better calibrate the LASP stellar parameters in future works. Similar calibration has already done for LSP3 \citep[LAMOST Stellar Parameter Pipeline at Peking University,][]{Xiang2015}. Compared to their work, we have a larger sample of star clusters except M67 which is in test plates and not included in the public LAMOST DR2 catalog.

For open clusters (whose metallicity is solor-like), there is a trend of decreasing estimated value with the increasing metallicity, namely, our estimated value are lower for metal-rich clusters and higher for metal-pool clusters than true value (from MWSC). A linear fit is applied on these 11 open clusters, the slope of $-0.57\pm0.13$ is found. This nonnegligible trend may be due to 1) our member star candidates are contaminated by a fraction of field star whose metallicity are solar-like or 2) the LASP pipeline underestimates metallicity for metal-rich stars and overestimates metallicity for metal-pool stars, i.e., the LASP tends to give a solar-like metallicity value for stars. However, at the extremely metal-pool end, the estimated metallicity is lower by 1$\sim$2\,$\sigma$.
With the limited star candidates it is difficult to give a definitive inference. But according to the large deviation of the slope from its error, it is likely that both of these two reasons contributes. To know exactly which is the main reason, more reliable member candidates and more star clusters are needed.

\section{Conclusions}\label{sect:summary}

In this work, we employ the method, combining the photometric with the kinematic information, to identify the member candidates of about 457 star clusters covered by LAMOST footprints. Although the LAMOST DR2 catalog contains about 3.8 million stellar spectra, only 24 star cluster are found member candidates more than 3 for each in the catalog. With these member candidates, for the first time, we are able to determine the median \rv\ and \feh\ for 4 and 11 clusters respectively. These member clusters are also very helpful in the assessment of the performance of the stellar parameters estimated from the LAMOST pipeline. Comparing the \rv\ with the MWSC catalog for the 24 clusters, we find the uncertainty of the velocity estimation in the LAMOST catalog is about \rvsig\,\kms, with about \rvmiu\,\kms\ offset from zero point. The uncertainty of \feh\ is also determined as \fehsig\,dex from the comparison with the MWSC catalog and we find the offset of \feh\ is \fehmiu\,dex. We notice that the LAMOST \feh\ is not consistent with values from the MWSC, implying some bias as a function of the true \feh. This is possibly due to both of the contributions from contamination of field stars in our member star candidates and the biased metallicity estimated by the LASP.

We propose to assign higher priority in the photometric member candidates for the clusters covered by the LAMOST footprints, so that they can be more possibly observed in the rest of the survey. In this way, the member candidates of the star clusters can be significantly increased. This will be very helpful in the study of the overall structure of the Milky Way as tracers as well as to better calibrate the stellar parameters, especially the \rv\ and \feh, in the LAMOST pipeline.

\begin{acknowledgements}
This work is supported by the Strategic Priority Research Program "The Emergence of Cosmological Structures" of the Chinese Academy of Sciences, Grant No. XDB09000000 and the National Key Basic Research Program of China 2014CB845700. CL acknowledges the National Science Foundation of China (NSFC) under grants 11373032, 11333003, and U1231119. XYC acknowledges the NSFC under grant 11403036.
The authors are grateful for constructive comments given by the referee.
Guoshoujing Telescope (the Large Sky Area Multi-Object Fiber Spectroscopic Telescope LAMOST) is a National Major Scientific Project built by the Chinese Academy of Sciences. Funding for the project has been provided by the National Development and Reform Commission. LAMOST is operated and managed by the National Astronomical Observatories, Chinese Academy of Sciences.
Facilities: LAMOST.

\end{acknowledgements}


\DeclareRobustCommand{\disambiguate}[3]{#1}
\bibliographystyle{raa}
\bibliography{bibtex}
\begin{appendix}
\section{}
\begin{sidewaystable}[anticlockwise]
\vspace{120mm}
\caption[]{Parameters of 24 star clusters.\label{tab:MWSC28p}}
\setlength{\tabcolsep}{1pt}
\tiny
\begin{tabular}{ lllcllcccl|lll|llll|llll|lll }
\hline
\hline
Name & RAJ2000 & DEJ2000 & Type & r0 & r2 & d & log$t$ & CMD & Fit Range(RV,RV$_z$) &RV & RV$^\ast$ & RV$_z^\ast$ &oRV$^\ast$   & Eff. oRV   & $\bar{P}_c^{RV}$   & Scale$_{RV}$ & oRV$_z^\ast$ & Eff. oRV$_z$ & $\bar{P}_c^{RV_z}$ & Scale$_{RV_z}$ & [Fe/H] & [Fe/H]$^\ast$ & Note \\
 & (deg) & (deg) & & (deg) & (deg) & (pc) & & & (kms$^{-1}$) & (kms$^{-1}$) & (kms$^{-1}$) & (kms$^{-1}$) & & & & & & & &  & (dex) & (dex) &   \\
\hline
$NGC 2632$ & $130.09$ & $19.69$ &  & $0.20$ & $3.10$ & $187$ & $8.92$ &y & $[15.0,50.0] [15.0,50.0]$ &$33.40 \pm 0.52$ & $30.59^{+5.39}_{-4.78}$ & $30.58^{+5.40}_{-4.80}$ &$341$ & $83$ & $0.24$ & $0.88$ &$408$ & $95$ & $0.23$ & $0.88$ &$0.09 \pm 0.11$ & $-0.06^{+0.26}_{-0.28}$    \\
$NGC 2168$ & $92.30$ & $24.36$ &  & $0.04$ & $0.98$ & $830$ & $8.26$ &y & $[-20.0,6.0] [-20.0,6.0]$ &$-10.80 \pm 1.73$ & $-9.56^{+3.61}_{-4.92}$ & $-9.59^{+3.60}_{-4.80}$ &$390$ & $82$ & $0.21$ & $0.98$ &$649$ & $56$ & $0.09$ & $0.98$ &$-0.16 \pm 0.09$ & $-0.11^{+0.16}_{-0.15}$    \\
$NGC 2099$ & $88.09$ & $32.57$ &  & $0.06$ & $0.48$ & $1400$ & $8.55$ &y & $[-7.0,17.0] [-7.0,17.0]$ &$7.70 \pm 0.02$ & $1.73^{+3.27}_{-2.83}$ & $1.80^{+3.16}_{-3.00}$ &$113$ & $16$ & $0.14$ & $0.96$ &$193$ & $27$ & $0.14$ & $0.96$ &$0.09 \pm 0.14$ & $-0.15^{+0.23}_{-0.20}$    \\
$NGC 1746$ & $76.14$ & $23.83$ &  & $0.05$ & $0.60$ & $800$ & $8.64$ &y & $[-7.0,16.0] [-7.0,16.0]$ &$2.00 \pm 3.70$ & $5.91^{+3.58}_{-3.52}$ & $6.00^{+3.52}_{-3.60}$ &$85$ & $6$ & $0.07$ & $1.05$ &$187$ & $10$ & $0.05$ & $1.05$ &$NaN \pm NaN$ & $-0.17^{+0.27}_{-0.29}$    \\
$NGC 1039$ & $40.55$ & $42.79$ &  & $0.16$ & $0.73$ & $510$ & $8.38$ &y & $[-25.0,6.0] [-25.0,6.0]$ &$-18.20 \pm 1.40$ & $-11.75^{+2.56}_{-3.97}$ & $-11.69^{+2.40}_{-4.06}$ &$113$ & $22$ & $0.20$ & $1.02$ &$161$ & $21$ & $0.13$ & $1.02$ &$0.07 \pm 0.03$ & $-0.06^{+0.18}_{-0.23}$    \\
$NGC 1647$ & $71.50$ & $19.17$ &  & $0.04$ & $0.75$ & $572$ & $8.30$ &y & $[-23.0,0.0] [-23.0,0.0]$ &$-6.10 \pm 0.39$ & $-10.86^{+3.59}_{-3.81}$ & $-10.79^{+3.60}_{-3.90}$ &$81$ & $19$ & $0.24$ & $1.03$ &$141$ & $33$ & $0.23$ & $1.03$ &$NaN \pm NaN$ & $-0.06^{+0.14}_{-0.28}$    \\
$NGC 2183$ & $92.70$ & $-6.20$ & n & $0.03$ & $0.39$ & $1047$ & $7.15$ &y & $[3.0,34.0] [3.0,34.0]$ &$10.30 \pm NaN$ & $19.66^{+4.38}_{-6.56}$ & $19.79^{+4.29}_{-6.60}$ &$108$ & $24$ & $0.23$ & $0.89$ &$123$ & $21$ & $0.17$ & $0.89$ &$NaN \pm NaN$ & $-0.09^{+0.16}_{-0.21}$    \\
$NGC 1912$ & $82.22$ & $35.80$ &  & $0.03$ & $0.39$ & $1144$ & $8.35$ &y & $[-20.0,24.0] [-20.0,24.0]$ &$1.00 \pm 0.58$ & $0.18^{+3.95}_{-4.11}$ & $0.30^{+3.90}_{-4.12}$ &$85$ & $30$ & $0.36$ & $0.92$ &$178$ & $49$ & $0.28$ & $0.92$ &$NaN \pm NaN$ & $-0.12^{+0.14}_{-0.17}$    \\
$NGC 2281$ & $102.08$ & $41.08$ &  & $0.04$ & $0.41$ & $500$ & $8.79$ &y & $[0.0,39.0] [0.0,39.0]$ &$13.30 \pm 4.11$ & $17.04^{+5.04}_{-5.16}$ & $17.09^{+5.10}_{-5.10}$ &$233$ & $139$ & $0.60$ & $0.79$ &$289$ & $150$ & $0.52$ & $0.79$ &$0.13 \pm 0.11$ & $-0.09^{+0.12}_{-0.18}$    \\
$NGC 2420$ & $114.60$ & $21.57$ &  & $0.02$ & $0.30$ & $2880$ & $9.37$ &y & $[47.0,87.0] [47.0,87.0]$ &$67.00 \pm 1.83$ & $67.90^{+6.02}_{-7.10}$ & $67.90^{+5.97}_{-7.05}$ &$74$ & $41$ & $0.55$ & $1.05$ &$87$ & $50$ & $0.58$ & $1.05$ &$-0.38 \pm 0.07$ & $-0.26^{+0.08}_{-0.08}$    \\
$NGC 2158$ & $91.86$ & $24.09$ &  & $0.04$ & $0.25$ & $4770$ & $9.33$ &y & $[9.0,45.0] [9.0,35.0]$ &$28.00 \pm 4.08$ & $33.41^{+3.19}_{-4.41}$ & $33.43^{+3.18}_{-4.53}$ &$16$ & $3$ & $0.18$ & $0.85$ &$45$ & $5$ & $0.12$ & $0.85$ &$-0.25 \pm 0.09$ & $-0.32^{+0.17}_{-0.23}$    \\
$NGC 869$ & $34.74$ & $57.15$ &  & $0.06$ & $0.54$ & $2300$ & $7.28$ &n & $[-46.0,-19.0] [-46.0,-19.0]$ &$-41.80 \pm 1.57$ & $-37.76^{+3.85}_{-3.91}$ & $-37.77^{+3.90}_{-3.90}$ &$57$ & $15$ & $0.26$ & $0.95$ &$254$ & $82$ & $0.32$ & $0.95$ &$-0.30 \pm NaN$ & $-0.12^{+0.11}_{-0.16}$    \\
$NGC 1662$ & $72.11$ & $10.93$ &  & $0.02$ & $0.48$ & $437$ & $8.70$ &n & $[-35.0,5.0] [-35.0,5.0]$ &$-13.50 \pm 0.40$ & $-16.89^{+3.74}_{-4.06}$ & $-16.94^{+3.75}_{-4.05}$ &$94$ & $30$ & $0.32$ & $1.00$ &$122$ & $37$ & $0.31$ & $1.00$ &$-0.10 \pm 0.01$ & $-0.17^{+0.20}_{-0.16}$    \\
$Basel 11B$ & $89.56$ & $21.99$ &  & $0.01$ & $0.18$ & $1318$ & $8.98$ &y & $[-2.0,26.0] [-4.0,25.0]$ &$NaN \pm NaN$ & $15.28^{+2.06}_{-5.14}$ & $15.29^{+2.10}_{-5.28}$ &$24$ & $5$ & $0.22$ & $0.88$ &$53$ & $13$ & $0.24$ & $0.88$ &$NaN \pm NaN$ & $-0.14^{+0.27}_{-0.23}$    \\
$NGC 1528$ & $63.85$ & $51.19$ &  & $0.05$ & $0.45$ & $950$ & $8.55$ &y & $[-44.0,6.0] [-44.0,6.0]$ &$-9.20 \pm 0.40$ & $-15.12^{+7.23}_{-9.46}$ & $-15.14^{+7.22}_{-9.44}$ &$106$ & $51$ & $0.48$ & $0.72$ &$230$ & $73$ & $0.32$ & $0.72$ &$NaN \pm NaN$ & $-0.08^{+0.27}_{-0.19}$    \\
$NGC 2252$ & $98.62$ & $5.37$ &  & $0.02$ & $0.22$ & $900$ & $8.86$ &y & $[6.0,32.0] [6.0,32.0]$ &$9.00 \pm 7.40$ & $16.82^{+4.37}_{-2.06}$ & $16.79^{+4.53}_{-2.13}$ &$23$ & $8$ & $0.34$ & $0.90$ &$22$ & $3$ & $0.14$ & $0.90$ &$NaN \pm NaN$ & $-0.07^{+0.32}_{-0.30}$    \\
$Basel 4$ & $87.25$ & $30.20$ &  & $0.02$ & $0.12$ & $2801$ & $8.66$ &y & $[-30.0,8.0] [-33.0,10.0]$ &$NaN \pm NaN$ & $-9.63^{+7.79}_{-6.16}$ & $-9.59^{+7.88}_{-6.10}$ &$29$ & $13$ & $0.46$ & $0.82$ &$63$ & $12$ & $0.19$ & $0.82$ &$NaN \pm NaN$ & $-0.23^{+0.14}_{-0.15}$    \\
$NGC 1960$ & $84.08$ & $34.16$ &  & $0.03$ & $0.26$ & $1200$ & $7.57$ &y & $[-11.0,37.0] [-11.0,37.0]$ &$-1.20 \pm 4.60$ & $9.43^{+10.59}_{-7.38}$ & $9.44^{+10.49}_{-7.37}$ &$36$ & $11$ & $0.30$ & $0.87$ &$90$ & $32$ & $0.35$ & $0.87$ &$NaN \pm NaN$ & $-0.20^{+0.23}_{-0.21}$    \\
$NGC 6205$ & $250.42$ & $36.46$ & g & $0.07$ & $0.47$ & $7107$ & $10.10$ &y & $[-270.0,-230.0] [-270.0,-230.0]$ &$-244.20 \pm 0.20$ & $-251.05^{+1.42}_{-1.38}$ & $-251.08^{+1.56}_{-1.29}$ &$4$ & $4$ & $1.00$ & $0.80$ &$8$ & $8$ & $0.95$ & $0.80$ &$-1.45 \pm NaN$ & $-1.67^{+0.09}_{-0.13}$    \\
$Waterloo 2$ & $82.01$ & $40.35$ &  & $0.01$ & $0.19$ & $550$ & $8.33$ &y & $[-39.0,15.0] [-39.0,15.0]$ &$NaN \pm NaN$ & $-7.92^{+11.62}_{-9.25}$ & $-7.79^{+11.39}_{-9.26}$ &$47$ & $17$ & $0.37$ & $0.71$ &$71$ & $17$ & $0.24$ & $0.71$ &$NaN \pm NaN$ & $-0.10^{+0.23}_{-0.19}$    \\
$NGC 6819$ & $295.32$ & $40.20$ &  & $0.02$ & $0.20$ & $2360$ & $9.21$ &n & $[-18.0,30.0] [-18.0,30.0]$ &$1.00 \pm 2.00$ & $0.95^{+8.18}_{-4.24}$ & $0.90^{+8.24}_{-4.20}$ &$80$ & $41$ & $0.51$ & $0.83$ &$109$ & $52$ & $0.48$ & $0.83$ &$0.09 \pm 0.03$ & $0.01^{+0.10}_{-0.28}$    \\
$NGC 5272$ & $205.55$ & $28.38$ & g & $0.07$ & $0.44$ & $10194$ & $10.10$ &n & $[-171.0,-138.0] [-200.0,-116.0]$ &$-147.60 \pm 0.20$ & $-154.06^{+3.97}_{-3.17}$ & $-153.94^{+3.75}_{-3.15}$ &$10$ & $8$ & $0.82$ & $0.80$ &$12$ & $10$ & $0.83$ & $0.80$ &$-1.34 \pm NaN$ & $-1.54^{+0.21}_{-0.24}$    \\
$Berkeley 71$ & $85.24$ & $32.27$ &  & $0.02$ & $0.14$ & $3260$ & $9.02$ &n & $[-48.0,-6.0] [-50.0,-8.0]$ &$NaN \pm NaN$ & $-25.50^{+5.30}_{-6.70}$ & $-25.48^{+5.28}_{-6.66}$ &$26$ & $16$ & $0.60$ & $0.67$ &$45$ & $25$ & $0.55$ & $0.67$ &$NaN \pm NaN$ & $-0.26^{+0.25}_{-0.34}$    \\
$Berkeley 32$ & $104.53$ & $6.43$ &  & $0.02$ & $0.12$ & $4996$ & $9.45$ &y & $[81.0,119.0] [81.0,119.0]$ &$101.00 \pm 3.30$ & $101.56^{+4.43}_{-9.31}$ & $101.48^{+4.59}_{-9.26}$ &$14$ & $12$ & $0.85$ & $2.00$ &$19$ & $16$ & $0.83$ & $2.00$ &$-0.29 \pm 0.04$ & $-0.33^{+0.19}_{-0.20}$    \\
\hline
\end{tabular}
\tablecomments{0.86\textwidth}{\\
\emph{Name} is the name of the cluster. \\
\emph{RAJ2000} and emph{DEJ2000} are the RA and Dec of the center of the cluster in degrees. \\
\emph{Type} is the type of the cluster whose value is null when it is an open cluster. And g denotes a globular cluster while n denotes open clusters with nebulosity.\\
\emph{r0} and \emph{r2} are the angular radius of the core and cluster respectively.\\
\emph{CMD} is set to be 'y' if CMD cut is applied and 'n' if not.\\
\emph{Fit Range} are parameters set during the selection of member star candidates using RV.\\
\emph{RV, RV$^\ast_z$} and \emph{RV$_z^\ast$} are the RV of the each star cluster estimated by MWSC, LAMOST \rv\ and \rvz.\\
\emph{oRV} is the number of member star candidates, and \emph{Eff. oRV} are the sum of the membership probability. $\bar{P}_c^RV$ are the mean membership probability, namely the \emph{Eff. oRV} / \emph{oRV}. \emph{Scales$_{rV}$} is the scale factor described in Sect~\ref{subsect:rv}. The corresponding quantities for member candidates of \rvz\ version are also presented.
}
\end{sidewaystable}

\setlength{\extrarowheight}{1.5pt}
\tiny
\begin{longtable}{ccccccccccc}
\caption{Member stars of 24 clusters and their parameters.}\label{tab:table_member_rv}\\
\hline \hline
Cluster & RA & Dec & $T_{\rm eff}$ & $T_{\rm eff}$ Err. &　log$g$　& log$g$ Err. & [Fe/H] & [Fe/H] Err. & RV & $P_c^{RV}$ \\
&(deg)&(deg) &(K)  &(K)  &(dex)  &(dex)  &(dex) &(dex)  &(kms$^{-1}$)  & (\%) \\
\hline
\endfirsthead
(Continued.)\\
\hline \hline
Cluster & RA & Dec & $T_{\rm eff}$ & $T_{eff}$ Err. &　log$g$　& log$g$ Err. & [Fe/H] & [Fe/H] Err. & RV & $P_c^{RV}$ \\
 &(deg)&(deg) &(K)  &(K)     &(dex)  &(dex)  &(dex) &(dex)  &(kms$^{-1}$)  & (\%) \\
\hline
\endhead
\hline
\endfoot
\hline
\endlastfoot
NGC 2632 & 134.269520 & 22.300831 & 5301.91 & 276.58 & 3.77 & 0.57 & -0.12 & 0.36 & 30.95 & 27.94\\
NGC 2632 & 134.496130 & 22.978804 & 6149.40 & 225.52 & 4.29 & 0.50 & -0.17 & 0.23 & 26.75 & 23.43\\
NGC 2632 & 133.440730 & 22.749496 & 5980.00 & 200.68 & 4.56 & 0.48 & 0.10 & 0.20 & 31.02 & 27.95\\
NGC 2632 & 132.641230 & 24.844702 & 5744.92 & 111.08 & 4.09 & 0.50 & 0.32 & 0.11 & 26.24 & 22.48\\
NGC 2632 & 133.188990 & 24.971348 & 5525.08 & 144.23 & 4.63 & 0.45 & -0.24 & 0.18 & 32.36 & 27.76\\
NGC 2632 & 133.079140 & 24.141962 & 5633.89 & 213.81 & 4.06 & 0.51 & 0.53 & 0.19 & 36.20 & 23.08\\
NGC 2632 & 132.775320 & 23.980831 & 4835.26 & 100.66 & 3.62 & 0.29 & 0.45 & 0.13 & 29.39 & 27.08\\
NGC 2632 & 133.595490 & 24.065157 & 6251.66 & 143.33 & 4.19 & 0.47 & 0.06 & 0.13 & 30.22 & 27.67\\
\hline
\end{longtable}

\end{appendix}

\end{document}